\documentclass[11pt]{article}

\DeclareMathAlphabet{\mathitbf}{T1}{cmr}{bx}{it}
\def\bbox#1{{ \mbox{\boldmath $#1$}}}

\newcommand 	\be   	{\begin{equation}}
\newcommand 	\ee   	{\end{equation}}

\newcommand {\ie}       {{\em i.e.,}\ }
\newcommand {\eg}       {{\em e.g.,}\ }
\newcommand {\etal}       {{\em et al.}\ }

\newcommand {\qEA}	{q_{\mathrm{EA}}}
\newcommand {\hp}	{\Delta h}
\newcommand {\tw}	{t_\mathrm{w}}
\newcommand {\Tc}	{T_\mathrm{c}}

\newcommand {\Tg}	{T_{\mathrm{g}}}
\newcommand{\kB}        {k_\mathrm{B}}

\usepackage{times}
\usepackage{graphicx} 
\usepackage{epsfig}

\begin{document}

\title{Low Temperature Properties of Finite Dimensional Ising Spin
Glasses: (some) Numerical Simulations}

\author{
Juan J. Ruiz-Lorenzo\\[0.6em] 
{\small  Departamento de F\'{\i}sica, Facultad de Ciencias, Universidad  de
  Extremadura.}\\[0.3em]
{\small Avda Elvas s/n. 06071 Badajoz. Spain.}\\[0.3em]
{\small and}\\[0.3em]
{\small  Instituto de Biocomputaci\'on y F\'{\i}sica de Sistemas 
Complejos.}\\[0.3em]
{\small Universidad de Zaragoza. 50009 Zaragoza, Spain.}\\[0.3em]
{\small ruiz@unex.es}\\[0.3em]
}

\date{21 April 2003}

\maketitle

\begin{abstract}

We review some recent results on finite dimensional spin glasses by
studying recent numerical simulations and their relationship with
experiments.  In particular we will show results obtained at zero and
non zero temperature, focusing in the low temperature properties of
the model, and contrast them with different pictures of the low
temperature phase of spin glasses: Replica Symmetry Breaking, Droplet
Model and Trivial-Not-Trivial Scenario.

\end{abstract}  

\section{Introduction}

Spin glasses are still a problematic issue. The introduction of
frustration and disorder in a statistical model has posed  a real
challenge to both experimentalist and theoreticians.

One can take, as an example, the ``canonical spin glass'': a metal
({\it e.g.}  Copper) in which ferromagnetic impurities have been
introduced ({\it e.g.}  Manganese). This system can be studied in the
RKKY framework and the result is an oscillating interaction which
couples the magnetic moments of the material. This oscillatory
behavior induced by the disorder (magnetic moments) also introduces
frustration in the material~\cite{EXPBOOK}.

In this work, given the limitations  of space, we have
restricted ourselves to treating a few topics related to numerical
simulations in finite dimensional spin glasses (only on Ising like
models), focusing on the properties of the low temperature phase, yet
we will treat them in detail. In the last years, a large amount of
work on finite temperature numerical simulations but also  zero
temperature ones and experiments have been done. We will try to
give a detailed description of some of these simulations and
experiments, highlighting common observables in them and contrasting
these results with some of the three main theoretical models: Droplet
Model (DM)~\cite{DROPLET}, TNT (Trivial overlap but Not Trivial link
overlap)~\cite{MARTIN,PAYO_2000} 
and Replica Symmetry Breaking (RSB)~\cite{PARISI,MEPAVI,RSB}. Related
work can be found in~\cite{NS}.

Unfortunately, we have put aside in this work interesting studies on
rheology~\cite{REO,STARIOLO},
ultrametricity~\cite{ULTRA,ULTRA_TH,FRART}, Heisenberg spin
glasses~\cite{Campbell,PY}, suitability of the Edwards-Anderson model
to describe real experiments~\cite{EA_moD}, two dimensional Ising spin
glasses~\cite{SG_2d,Middleton},
heterogeneity~\cite{CASTILLO,hete,Giorgio_hete,B_2003}, Sherrington-Kirkpatrick
model~\cite{Barbara1,Barbara2}, sum rules~\cite{RSB},
anisotropy~\cite{VMM},
chaos~\cite{RIZZO,Billoire,chaos_recent,chaos_droplet}, eigenvalues
analysis~\cite{CORRE}, three dimensional ferromagnetic spin
glasses~\cite{MARTIN3} and in field numerical
simulations~\cite{MARTIN1,MARTIN2,MARTIN4,SUE3}. The list of references given
in this paragraph is not complete.

Very good reviews and books have been written in the past years. We
refer the reader to them~\cite{BINYOU,FISHER,MEPAVI,EXPBOOK}. In
addition numerical simulations have been reviewed
in~\cite{BOOK,RIEGER}, experiments in~\cite{EXP} and dynamics
in~\cite{BOOKD,Bouchaud_2000,LETICIA,FELIX}.

To put  this work into context we will review (briefly) the three main
theoretical approaches to spin glasses.

The first one is the so-called Replica Symmetry Breaking. It is based
in the standard procedure which has worked extremely well in
Statistical Mechanics in the past decades (the paradigm is the ordered
Ising model). Firstly, one must solve  the model in the Mean Field
approximation. This is equivalent to solve the infinite
dimensional model exactly. This was done by G. Parisi in
1980~\cite{PARISI,MEPAVI}. His main results are that there exist a
(countable) infinite number of (finite volume) pure states organized
in an ultrametric fashion. The differences in extensive free energy
among these pure states are of order one. In addition, the interface
between two of these states is space filling (its surface scales as
its volume, like an sponge)~\cite{RSB}. The Parisi solution also
predicts a transition in magnetic field.

Once we know which is the solution in infinite dimensions (where there
are no fluctuations) we enable the system to fluctuate around the Mean
Field solution (in this case, that of Parisi). The appropriate
technique to handle this kind of problem is the Renormalization group
(that can be implemented in the Field Theoretical approach) and the
goals are computing the upper critical dimensions (above  which Mean
Field provides a good picture of the transition) and determining the
critical exponents (at fixed dimension or in the
$\epsilon$-expansion) below the upper critical dimension. Within this
approach it is very difficult (since it is based mainly in
perturbation theory) to estimate the lower critical dimension (the
largest dimension below which  there is not phase transition). 
The renormalization group program
has been done (in part) by de Dominicis, Temesvari and
Kondor~\cite{deDOMINICIS}. We should remark that this approach does
not change the low temperature properties. Hence in between the lower
critical and infinite dimensions the qualitative description of the broken
phase is still provided by the Parisi solution.

Another compelling theory is the droplet model~\cite{DROPLET}. 
The rationale of this
model is the Migdal-Kadanoff renormalization group. This technique is
exact in one dimension and is approximate in higher dimensions. The
main results of the DM is that there are two pure states (only one, if we
consider the global spin flip symmetry), and that the magnetic
field destroys the phase transition. In addition we can mention that the
typical excitations are compact domains of reversed (against the
ground state) spins. The cost in energy of these excitations
scales as a power of the typical size of the droplet, $L^\theta$.

Recently  has been proposed a third way which interpoles between
the droplet model and RSB: the TNT proposal~\cite{MARTIN,PAYO_2000}. 
In RSB $\theta=0$, since
we can create an excited state with ${\cal O}(1)$ energy. In this third
approach $\theta=0$ as in RSB but the link overlap is trivial (as in
the droplet model: the probability distribution of the link overlap is
delta peaked). In RSB the link overlap is believed to be proportional
to the squared of the overlap (in infinite dimension the link overlap
$q_l=q^2$, where $q$ is the overlap): as far as the probability distribution
of the overlap is not trivial then the probability distribution of the
link overlap must not be delta peaked. We refer the reader to the text
below for more details about the link overlap.

The chapter is organized as follows. In the first section we examine the
issue of the phase transition, giving strong numerical results which
support a finite transition at non zero temperature. Next we will
study the properties of the low temperature region (below the critical
point). In this part we show numerical simulations which highlight
physical properties which can be described consistently assuming a RSB
phase. Moreover we will describe the experimental computation of the
dynamical correlation length and the possible interpretations of the
different scalings proposed. In section 4 we will study the
generalization of the fluctuation dissipation theorem out of
equilibrium, starting with the analytical basis and continuing with
some numerical results which support the link statics-dynamics. This
tool is very important because it can be implemented in experiments (we
will show these). In section 5 we will show the memory/rejuvenations
experiments. In the following section we will study zero temperature
properties which probe the different theoretical pictures. Finally we
will return to non zero temperature and describe recent numerical
simulations computing the link overlap at finite temperature.

\section{On the phase transition}

This part of the review is devoted to showing numerical evidences which
favor strongly a phase transition in the three dimensional Ising spin
glass at finite temperature.

The existence of a phase transition in the three dimensional Ising
spin glass has been attacked mainly using finite size scaling (FSS)
methods~\cite{FSS}. In these methods one monitors which is the behavior
of some (critical) observables of the system when one changes the size
of it.  We will describe in this section how to implementate the FSS
to spin glasses and then how to define a good cumulant which signs
clearly the transition point.

The initial  point is to introduce the Edwards-Anderson Hamiltonian~\cite{EA}
\be
{\cal H}=-\sum_{<i,j>} J_{ij} \sigma_i \sigma_j \;, 
\ee 
where the sum
is extended to all the pairs of nearest neighbors, $\sigma_i=\pm 1$
are Ising variables and $J_{ij}$ are random (quenched) variables. In general
the $J_{ij}$ are  drawn for a Gaussian distribution with zero mean and
unit variance. One can also choose the random couplings from a bimodal 
distribution: $J_{ij} =\pm 1$ with equal probability.

It is well known that observables in spin-glasses need to be defined
in terms of real replicas, that is, for every disorder realization,
one considers two thermally independent copies of the system
$\{\sigma_i,\tau_i\}$~\cite{BOOK}. Observables are most easily defined
in terms of a spin-like field, the so-called overlap field (which is the order
parameter in spin glasses):
\begin{equation}
q_i= \sigma_i \tau_i\;.
\end{equation}
The total overlap is the lattice average of the $q_i$
\begin{equation}
q=\frac{1}{V} \sum_i q_i\;,
\end{equation}
while the (non-connected) spin-glass susceptibility is\footnote{As
  usual we use the brackets to denote the thermal average for a given
  choice of disorder, and the overline to mark the average over the disorder.}
\begin{equation}
\chi_q= V \overline{\langle q^2 \rangle }\; .
\label{chi}
\end{equation}
In Finite-Size Scaling studies, it is useful to have dimensionless
quantities, that go to a constant
value at the critical temperature.  The standard example of this 
quantity is the Binder cumulant
\begin{equation}
g_4=\frac{3}{2}- 
\frac{1}{2}
\frac{\overline{\langle q^4  \rangle }} 
{\overline{\langle q^2  \rangle}^2 } \; .
\label{binder}
\end{equation}
Another example is the $g_2$ cumulant~\cite{DISORDERED}, that measures
the lack of self-averageness of the spin-glass susceptibility
\begin{equation}
g_2=\frac{\overline{\langle q^2\rangle^2}-\overline{\langle q^2\rangle}^2}
{\overline{\langle q^2\rangle}^2}\;.
\end{equation}
In reference~\cite{MAFE} a third cumulant was proposed 
which is a function of $g_4$ and
$g_2$
\be
G=\frac{1}{2} \frac{g_2}{1-g_4} \;.
\ee

\begin{figure} 
\begin{center} 
\includegraphics[width=0.5\textwidth,height=0.3\textheight,angle=90]{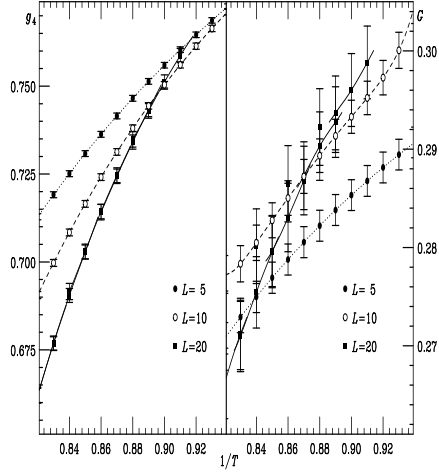}
\caption{$g_4$ and $G$ cumulants as a function of the inverse temperature
  for the three dimensional $\pm J$ Ising spin glass ~\cite{SUE1}.}
\label{fig:binder}
\end{center} 
\end{figure}

These  cumulants have been really useful to characterize phase
transitions both in ordered systems (the Binder cumulant) and in
disordered ones ($g_2$ in diluted Ising models). Nevertheless in Ising
spin glasses they do not provide a clear signature of the phase
transition (\ie a clear crossing between curves corresponding to
different lattice sizes). In Figure \ref{fig:binder} we show both
cumulants as a function of the temperature for the three dimensional
Ising model (with a binomial distribution for the couplings and
helicoidal boundary conditions).

Unfortunately, $g_4$ and $g_2$ require the evaluation of a four-point
correlation function, which is statistically a much noisier quantity
than a two-point one.  A more convenient observable is the
correlation-length, which is defined only in terms of the two-point
correlation function. Notice that its ratio with the lattice size
is again dimensionless~\cite{FSS}. We
therefore are faced with the problem of defining a correlation-length
on a finite lattice.  This was done in Ref.~\cite{COOPER}. The main
steps of the constructions are the following.  Let $C(\mathitbf{r})$
be the correlation function of the overlap field,
\begin{equation}
C(\mathitbf{r})=\frac{1}{V}\sum_i \overline{\langle q_i
q_{i+\mathitbf{r}}\rangle} 
\end{equation} 
and $\hat C(\mathitbf{k})$ its Fourier transform.  Notice that $\hat
C(0)$ is the
spin glass susceptibility.  Then, inside the
critical region on the paramagnetic side and in the thermodynamical
limit, one has
\begin{eqnarray}
\hat C(\mathitbf{k})&\propto&\frac{1}{k^2+\xi^{-2}}\ ,\quad 
\Vert{\mathitbf{k}}\Vert\ll \xi^{-1}\,,\\
\xi^{-2}&=&-\frac{1}{\hat C}\left.
\frac{\partial\hat C}{\partial \mathitbf{k}^2}
\right|_{\mathitbf{k}^2=0}\label{XIDERIVADA}\;.
\end{eqnarray}

On a finite lattice, the momentum is discretized, and one
uses~\cite{COOPER} a finite-differences approximation to
eq.~(\ref{XIDERIVADA}),
\begin{equation}
\xi^2=\frac{1}{ 4  \left[ \sin^2(k^x_{\mathrm m}/2)+\sin^2(k^y_{\mathrm m}/2)
+\sin^2(k^z_{\mathrm m}/2)\right]} 
\left[ \frac{\chi_q}{\hat C(\mathitbf{k}_{\mathrm m})}-1 \right] \; ,
\label{correlation}
\end{equation}
where $\chi_q$ was defined in eq.~(\ref{chi}) and 
$\mathitbf{k}_{\mathrm m}$ is the minimum wave-vector allowed 
for the 
boundary conditions used (\eg $\mathitbf{k}_{\mathrm
m}=(2\pi/L, 2 \pi /L^2, 2\pi/L^3)$ for helicoidal boundary
conditions). 
Of course, eq.~(\ref{XIDERIVADA})
holds in the thermodynamic limit ($L\gg \xi$) of the paramagnetic
phase. As we do not use connected correlation functions, $\xi$ has
sense as a correlation length only for\footnote{
  We use $\Tc$ to denote the critical temperature obtained in 
  numerical simulations and in theoretical computations 
  and $\Tg$ the one obtained in experiments.} $T>\Tc$.

We can study the scaling behavior of the finite-lattice
definition~(\ref{correlation}) on a critical point, where the
correlation function decays (in $d$ dimensions) as
$r^{-(d-2+\eta)}$. The behavior of the Fourier transform of the
correlation function for large $L$ in three dimensions is given by
\begin{equation}
\hat C(k)\sim \int_0^L d r\, r^{1-\eta}\frac{\sin(k r)}{k r}
\end{equation}
and one finds that $\chi_q/\hat C(\mathitbf{k}_{\mathrm m})$ goes to
a constant value, larger than unity, because 
$\Vert{\mathitbf{k}}_{\mathrm m}\Vert={\cal O}(1/L)$.  
Furthermore, $\xi/L$ tends to a
universal constant at a critical point (like the Binder cumulant
$g_4$).  Moreover, on a broken-symmetry phase, where the fluctuations
of the order parameter are not critical, one has $\chi_q={\cal O}(L^d)$,
while $\hat C(\mathitbf{k}_{\mathrm m})={\cal O}(1)$. Therefore the
full description of the scaling behavior of $\xi/L$ is as follows.
Let $\xi_\infty$ be the correlation-length in the infinite lattice: in
the paramagnetic phase, for $L\gg \xi_\infty$, one has $\xi/L={\cal
O}(1/L)$. In the scaling region, where $\xi_\infty\geq L$, $\xi/L={\cal
O}(1)$, while in a broken-symmetry phase on a lattice larger than the
scale of the fluctuations, $\xi/L={\cal O}(L^{d/2})$. Consequently, 
if one plots $\xi/L$ for several lattice sizes
as a function of temperature, the different graphs will cross at the
critical one.

\begin{figure} 
\begin{center} 
\includegraphics[width=0.5\textwidth,
height=0.3\textheight,angle=90]{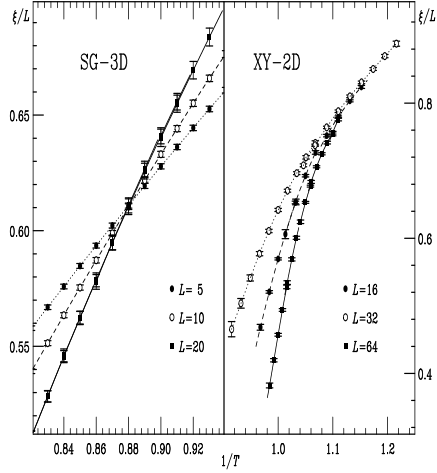}
\caption{$\xi/L$ cumulant for the three dimensional $\pm J$ Ising spin
  glass (left part of the figure). The same plot for the two
  dimensional XY (ordered) model. Taken from reference ~\cite{SUE1}.}
\label{fig:xi}
\end{center} 
\end{figure}

We can see the (clear) crossing phenomena in Figure \ref{fig:xi}. Also shown in 
this figure (right part) the same observable for the
two dimensional XY model (with no disorder). This double plot tells us
that 1) there is a phase transition a finite temperature and 2) we
should discard $\Tc=0$ and XY-like scenarios for the phase transition
of the three dimensional Ising spin glass~\footnote{These numerical
  results have been obtained with the dedicated computer
  SUE~\cite{SUE}, which has a performance of 0.2 ns/spin.}.

\section{Some properties of the low temperature region}

In this section we will describe numerical simulations and experiments
which try to discern which are the low temperature properties of the
three dimensional Ising spin glass by working well below the
transition point. We will start by discussing the properties at the
upper critical dimension of the model (which is six) where the
analytical predictions from RSB simplify. Then we will report
results in three dimensions. Finally we will review some issues
related to the behavior of the dynamical correlation length.

\subsection{$\bbox{d=6}$}

First, we will check one of the RSB predictions. To do this, numerical work
has been carried out  just at the upper critical dimension. In this dimension
there is no renormalization of the powers of propagators (\ie the
anomalous dimension vanishes) and only multiplicative factors
occur\footnote{At the upper critical dimension logarithmic corrections
appear. These have been studied numerically by Wang and
Young~\cite{WAYO}, and subsequently computed analytically in
reference~\cite{JJRL}. For a discussion on the lower critical dimension 
see reference~\cite{HART1}}. 

If RSB holds, a $1/p^4$ propagator should be found by looking at the
$q=0$ sector of the model in the broken phase ($T <
\Tc$)\footnote{This ergodic sector is very important. Out of
equilibrium, the system remains in this sector.}.
We remark that at
six dimensions the equilibrium overlap-overlap correlation
function constraint to $q=0$ was obtained by De Dominicis
\etal~\cite{deDOMINICIS} 
\be C_{\rm RSB}(x)|_{q=0} \sim \left\{
\begin{array}{lcl } x^{-4} & \mathrm{ if }\;& T=\Tc\;,\\ x^{-2} &
\mathrm{ if }\;& T<\Tc\;,
		      \end{array}
                 \right.
\label{eq:chi_rsb}
\ee 
which corresponds to $1/p^2$ at $T=\Tc$ (the usual critical propagator) 
and $1/p^4$ (the replicon mode) for $T<\Tc$.

From this correlation function 
we can compute the associated (spin glass) susceptibility
\be
\chi=\int d^6 x\; C(x) \;.
\label{chi_def}
\ee 
Since we are working on a finite lattice, the previous integral must 
be performed in a box of size $L$. If
we want to observe the dynamical behavior of $\chi$, the upper limit
in the integral should be changed to $\xi(t)$, the dynamical
correlation length. At this point we can assume that $\xi(t) \simeq
t^{1/z(T)}$, which defines an, in principle, effective dynamical
critical exponent, $z(T)$. Furthermore, one can  assume a functional dependence
$z(T)=4 \Tc/T$, where $\Tc$ is the critical temperature: with this temperature
dependence we recover the value $z$ at the critical temperature
$z(T_c)=4$ as predicted by Mean Field~\cite{MEPAVI}. 
The result for $\chi(t)$ is 
\be
\chi(t) \sim \left\{ \begin{array}{ll} t^{1/2} & \mathrm{ if}~ T=\Tc\;,\\
t^{4/z(T)} & \mathrm{ if}~ T<\Tc \;.
		      \end{array}
                 \right.
\ee
This expression can be compactly  written as
\be
\chi(t) \sim t^{h(T)} \;.
\ee
This formula should be   valid if we remain all the time in the $q=0$
sector. Hence, the exponent $h(T)$ is a discontinuous
function of temperature: \ie $h(\Tc^-)=1$ while
$h(\Tc^+)=1/2$. Moreover $h(T)$, if the Ansatz for $z(T)$ is right,
should grow linear.

This can be tested by performing an out of equilibrium numerical
simulation in a large lattice. The run starts at random and suddenly
the system is quenched below the critical temperature. At this point
the growth of the non linear susceptibility is recorded. At the same
time, one can check that the system (due to the large lattice simulated)
develops no overlap (and so we are sure that we are simulating inside
the $q=0$ sector of the theory).  The strategy is to point out the
discontinuity of the power of the $q=0$ propagator when we reach the
critical temperature from below (the propagator changes from $1/p^4$
to the standard and critical $1/p^2$ propagator). So, one needs to redo the
previous schedule but quenching to the critical temperature.

\begin{figure} 
\begin{center} 
\includegraphics[width=0.5\textwidth,height=0.3\textheight]{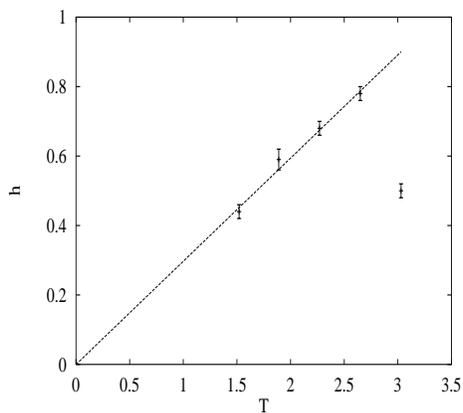} 
\caption{Susceptibility exponent as a function of the temperature in
  the six dimensional Ising spin glass~\cite{PRRTRL_1997}. Notice the
  linear region below the critical temperature and the discontinuity
  at the critical point.}
\label{6DIM}
\end{center} 
\end{figure}

In Figure \ref{6DIM} we plot the results and it is clear that the
system behaves as RSB predicts: $h(T)$ grows linear below the critical
point and develops a discontinuity, just on the amount predicted by
RSB, at the critical temperature. And so, it has been shown 1) the
existence of the replicon mode at finite dimensions and 2) the growth
of the correlation length can be described with the following law:
$\xi(T,t) \simeq t^{1/z(T)}$ with $z(T) \propto
1/T$~\cite{PRRTRL_1997}.

The next step is  trying to see if this picture holds in lower
dimensions, in particular in the physical dimension three.

\subsection{$\bbox{d=3}$}

In three dimensions, it is possible to 
 handle this problem (replicon mode in addition to
a given behavior of $\xi(t,T)$) by studying the decay with time and
position of the overlap-overlap correlation function,
\begin{equation}
C(x,t)= \frac{1}{L^3} \sum_i\overline{\langle \sigma_{i+x} \tau_{i+x}
\sigma_i \tau_i \rangle_t} \;.
\end{equation}
where $\sigma$ and $\tau$ are two real replicas (which evolve with the
same disorder) and the index $i$ runs over all the points of the
lattice.  As usual we denote by $\overline{(\cdot \cdot \cdot)}$ the
average over the disorder and, in this context, $\langle (\cdot \cdot
\cdot) \rangle_t$ is the average over the dynamical process (for a
given realization of the disorder) at time $t$. In plain words, the
two replicas ($\sigma$ and $\tau$) evolve with the same disorder but
with different random numbers.

\begin{figure} 
\begin{center} 
\includegraphics[width=0.5\textwidth,height=0.3\textheight]{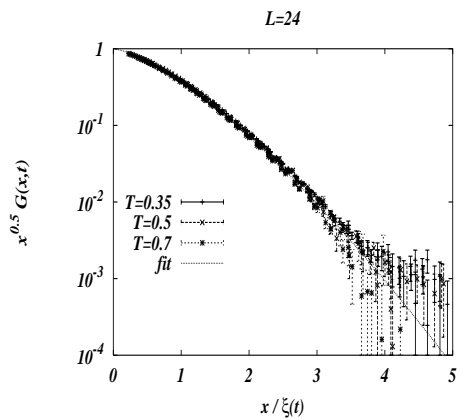} 
\caption{Re-scaled correlation function, $x^\alpha C(x,t)$, against
the scaling variable, $x/t^{1/z}$, for $L=24$ and $T=0.35, 0.5$ and
0.7. Taken from reference~\cite{MPRTRL_1999}. Notice the quality of
the scaling.}
\label{fig:scaling_corr}
\end{center} 
\end{figure}

In the $q=0$ sector (obtained simulating very large lattices, for large
times, but by controlling that the overlap of the system is always very
small) it has been  obtained that the numerical data~\cite{COR2,MPRTRL_1999} 
follow very well the
following scaling law (it  has been checked that this behavior also holds
in four dimensions~\cite{PRTRL4d,BB_2002}) 
\begin{equation}
C(x,t)=\frac{1}{x^\alpha}
\exp \left[ - \left( \frac{x}{\xi(t)} \right)^\delta  \right] \; .
\label{eq:ansatz_corr}
\end{equation}

We show in Figure \ref{fig:scaling_corr} the scaling plot for three
different temperatures and the fit using
eq.~(\ref{eq:ansatz_corr}). The scaling plot and the agreement with
the fit is very good. We can cite that the $\alpha$ exponent does not
show a clear temperature dependence in three dimension ($\alpha \simeq
0.5$)~\cite{COR2,MPRTRL_1999}\footnote{In three
  dimensions~\cite{MP_PRL} it has been found at zero temperature that
  $\alpha\simeq 0.4$, in good agreement with the value 
found at non zero temperature.}
 , whereas in four dimension the
situation is very different since the alpha exponent varies greatly
with temperature~\cite{PRTRL4d,BB_2002}.

\begin{figure} 
\begin{center} 
\includegraphics[width=0.5\textwidth,height=0.3\textheight,angle=0]
{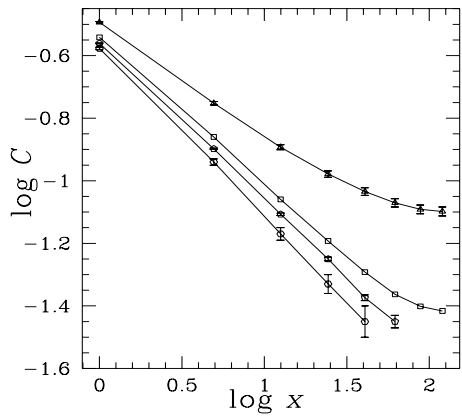}
\caption{We show four overlap-overlap correlations functions at
  $T=0.7$. From top to bottom: 1) equilibrium correlation function, 2)
  equilibrium 
(but computed with a small cut-off on the overlap) correlation function 3) and 4)
  extrapolation to infinite time of two dynamical correlation
  functions computed using two different annealing procedures. See the
  text for more details. Figure taken from reference~\cite{COR1}.}
\label{fig:cor_eq}
\end{center} 
\end{figure}

This scaling law provides us with the equilibrium form of the
propagator, by taking the limit  $t\to \infty$ in eq. (\ref{eq:ansatz_corr}):
\begin{equation}
C_\mathrm{eq}(x)\equiv \lim_{t\to \infty} C(x,t)\propto\frac{1}{x^\alpha} \;,
\label{eq:pure_law}
\end{equation}
where the proportionally constant is $C_\mathrm{eq}(x=1)$.
Of course, the exponent is not as found in six dimensions  due to the
renomalization effects (see eq. (\ref{eq:chi_rsb})). 

A further test can be done in three dimensions in order to check the
pure power law of the correlation function restricted to small
overlaps, eq.~(\ref{eq:pure_law}).  One can compare this behavior
(obtained dynamically and using an extrapolation) with that obtained
by computing at equilibrium the overlap-overlap correlation function
by taking only those measures with overlap $q<0.01$. 
In Figure \ref{fig:cor_eq} we plot in
the lower part of the figure two curves corresponding to the
correlation function obtained in a dynamical process taking the
extrapolation to infinite time. The upper curve is the equilibrium
correlation function (computed without imposing cut-off) and finally
the last curve is the equilibrium one computed using a small
cutoff ($q_\mathrm{max}=0.01$). 
The agreement between the lower three curves is really
good. This plot provides an additional evidence to the existence of a
replicon mode in three dimensions~\cite{COR1}.  In the droplet model
$C_\mathrm{eq}(x) \to \qEA^2$ as $x\to \infty$ in contrast with the
numerical results which support eq. (\ref{eq:pure_law}).

\subsection{Dynamical correlation length}

We have seen that in six dimensions 
the correlation length can be fitted $\xi(T,t)$ as
$t^{1/z(T)}$ with $z(T)=z_\mathrm{c} \Tc/T$, where $z_\mathrm{c}$ is
the dynamical critical exponent at the critical point ($\Tc$). In
particular, it was found in three dimensions
that~\cite{COR2,MPRTRL_1999,Kisker,RIEGER}
\be 
\xi(t,T) \propto t^{0.153(12) T/\Tc}\;,
\label{eq:xi}
\ee 
where we have assumed that $\Tc=0.95(3)$. In four
dimensions~\cite{PRTRL4d} a similar behavior was found 
\be 
\xi(t,T) \propto t^{0.19(1) T/\Tc}\;,
\ee 
where $T_c =1.80(1)$.  The behavior in four dimensions interpolates
very well between the three dimensional results and that obtained in
six dimensions $\xi(t,T) \propto t^{0.25 T/\Tc}$.

\begin{figure} 
\begin{center} 
\includegraphics[width=0.5\textwidth,height=0.3\textheight]{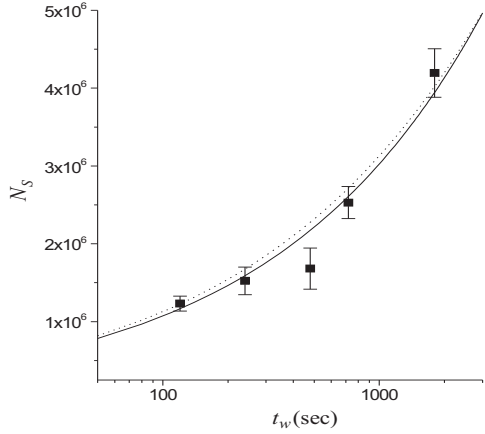} 
\caption{$N_s$, number of spins participating in barrier quenching
(and hopping) as a function of $\log \tw$ at $T=0.78 \Tg=28~
\mathrm{K}$ for $\mathrm{Cu}\underline{\mathrm{Mn}}$. 
The solid curve is the prediction
for power law dynamics, while the dashed one comes from activated
dynamics, see the text for more details. Taken from
reference~\cite{JOWHV_1999}.}
\label{fig:orbach}
\end{center} 
\end{figure}

This dependence of the dynamical correlation length with temperature
and time has been checked experimentally. The basic idea of the
experiment reported in reference~\cite{JOWHV_1999}, was to introduce
an external magnetic field and then operationally define the dynamical
correlation length via the volume of the droplet which contributes to
the Zeeman energy: $E_\mathrm{Zeeman} \propto N_\mathrm{s}\chi H^2$
(where $N_\mathrm{s}$ is the number of spins contributing to the
Zeeman energy, $H$ is the magnetic field and $\chi$ is the magnetic
susceptibility).  By effect of the magnetic field, the typical times
of the dynamics are modified by a factor $\exp(-c N_\mathrm{s} \chi
H^2/T)$, where $c$ is a numerical factor. By measuring this reduction
factor one can extract the number of spins involved in the dynamics
for a given waiting time and temperature and using that $N_\mathrm{s}
\propto \xi(\tw,T)^3$, $\xi(\tw,T)$ can be computed.  In this way they
computed the correlation length and by performing the experiment at
different temperatures. The following experimental
dependence was found (see the solid line in Figure \ref{fig:orbach}): \be
\xi(\tw,T) = 0.653 \left(\frac{\tw}{\tau_0} \right)^{0.169 T/\Tg} \;,
\ee where $1/\tau_0=4.1\times 10^{12} ~\mathrm{s}^{-1}$.  The
agreement with the result obtained in numerical simulations (see
eq. (\ref{eq:xi})) is very good. Nevertheless a fit assuming activated
dynamics (droplet model) is also possible (see the dashed line in
Figure \ref{fig:orbach}), obtaining \be \xi(t,T) = 10^{-5}
\left[\frac{T}{\Tg} \log\left(\frac{\tw}{\tau_0}\right)
\right]^\frac{1}{0.21} \;.
\label{eq:act}
\ee However we see that the prefactor of the fit is really small (it
would be natural for it to be ${\cal O}(1)$). Moreover the $\psi$ exponent
is just at the lowest allowable value in the droplet model
($\psi=0.2$). However, numerical work suggests that
$1/\psi=1/0.7$~\cite{Kisker,RIEGER}.

A plausibility argument for the linear dependence of the effective dynamical
critical exponent $z(T)$ with the inverse of the temperature was given
by H. Rieger in reference~\cite{RIEGER}.  Assuming an Arrhenius law
and that free energy barrier for an excitation of typical size $L$,
scales as $\log L$ (\ie $\psi=0$), we can obtain
that\footnote{We introduce in this discussion the Boltzmann constant,
$\kB$, which has been set to one in the rest of the paper.}  
\be
\tau \propto \exp\left(c \frac{\log L}{\kB T}\right)\; , 
\ee 
where $c$ is a constant, which can be rewritten as $\tau \propto L ^{1/z(T)}$ with
$z(T)\propto 1/T$.

However, a different argument based in the droplet picture and also
accounting for the experimental data can be given~\cite{BDHV_2001}.
Indeed, let us assume that the time needed to evolve a conformation on
a scale of size $l_n$ is given by (this defines the $\psi$ droplet
exponent)\footnote{By inverting in this formula $l_n$ in terms of
$t_n$ we obtain the activated dynamics prediction for the dynamical
correlation length, see equation (\ref{eq:act}).}  
\be 
t_n=t(l_n) \sim
\tau_0 \exp\left(\frac{\Upsilon l_n^\psi}{\kB T}\right) \;.  
\ee 
This behavior has been tested in Figure \ref{fig:orbach} and although the  the fit is
good the parameters are not realistic enough (see above). Nonetheless
it is possible to modify the previous formula in order to work in the
neighborhood of the phase transition 
\be t_n=t(l_n) \sim \tau_0
l_n^{z_\mathrm{c}} \exp\left(\frac{\Upsilon(T) l_n^\psi} {\kB
T}\right) \;,
\label{eq:bouchaud}
\ee
with $\Upsilon(T)=\Upsilon_0 (\Tc-T)^{\nu \psi}$. Near the phase
transition this formula reduces to the usual (non activated) formula
$\tau \simeq l_n ^{z_\mathrm{c}}$.

\begin{figure} 
\begin{center} 
\includegraphics[width=0.5\textwidth,height=0.3\textheight,angle=270]{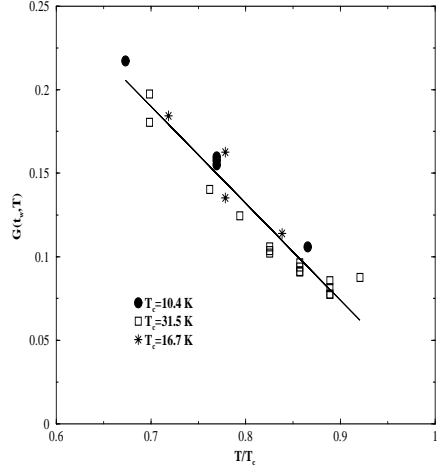} 
\caption{Plot of $G(\tw,T)$, defined in the text, against $T/\Tg$ for
  different waiting times, temperatures and three spin glasses with
  different critical temperatures $\Tg$. The authors have used
  $\psi=1.5$, $\nu=1.3$ and $z_\mathrm{c}=5$ motivated by the
  experimental study of   $\mathrm{Ag}\underline{\mathrm{Mn}}$. If
  $z_\mathrm{c}=6$ is assumed then the data extrapolate to
  $T/\Tg=1$. The scaling is very good. From reference~\cite{BDHV_2001}.}
\label{fig:bouchaud}
\end{center} 
\end{figure}

To test this generalization of the original droplet formula it is
interesting to compute experimentally the following function (using the
same procedure as in ref.~\cite{JOWHV_1999}):
\be
G(\tw,T)=\left( \frac{\log(\tw/\tau_0)-\frac{z_\mathrm{c}}{3} \log
  N_\mathrm{s}(\tw,T)}
{\frac{\Tg}{T} N_\mathrm{s}(\tw,T)^{\psi/3}} \right)^{\frac{1}{\nu \psi}} \;.
\ee
In Figure \ref{fig:bouchaud}, $G(\tw,T)$ is shown  against  $T/\Tg$ for
different waiting times, temperatures and three different spin glasses
with different critical temperatures. The linear fit, supporting
equation (\ref{eq:bouchaud}),  is very good and the points extrapolate
to near 1 when $G$ approaches zero.

In addition, Berthier and Bouchaud, in ref.~\cite{BB_2002}, 
have tested this scenario via numerical
simulations. In particular in four dimensions they have found that
this droplet generalization works well. However the microscopic time
they obtained in their fits shows (in three dimensions) a non
monotonic dependence on the temperature, for which there is not 
physical explanation~\cite{BB_2002}.

\section{Off equilibrium fluctuation-dissipation relations}

One of the most important results of Statistical Physics at
equilibrium is the so-called fluctuation-dissipation theorem. In this
section we will review its theoretical basis and its generalization at
early times in the dynamics. Moreover we will see how this
generalization provides us with a useful tool to understand which are
the properties of the low temperature phase at equilibrium.

\subsection{Theoretical basis}

The starting point is to  perturb the original Hamiltonian, $\cal{H}$, of a spin
glass in a magnetic field as 
\begin{equation}
{\cal H}^\prime= {\cal H} + \int \hp(t) A(t)\,d t \;, 
\end{equation}
where
\be
{\cal H}=-\sum_{<i,j>} J_{ij} \sigma_i \sigma_j +h \sum_i \sigma_i \;, 
\ee 
$h$ being the magnetic field. 
We can define 
the following
autocorrelation function
\begin{equation}
C(t_1,t_2) \equiv \langle A(t_1) A(t_2) \rangle \;. 
\label{eq:auto}
\end{equation}
Usually, $A(t)=\sigma_i(t)$,
and the associated response function
\begin{equation}
R(t_1,t_2) \equiv \left. \frac{\delta \langle A(t_1) \rangle}{\delta
\hp(t_2)}\right|_{\hp=0} \; .
\label{res}
\end{equation}
The brackets $\langle \cdots \rangle$ in eq.~(\ref{eq:auto}) and
eq.~(\ref{res}) imply here a double average, one over the dynamical
process and one over the disorder.

In the dynamical framework, assuming time translational invariance, it
is possible to derive the fluctuation-dissipation theorem (FDT), that reads
\begin{equation}
R(t_1,t_2)=\beta \theta(t_1-t_2) \frac{\partial
C(t_1,t_2)}{\partial t_2} \; ,
\protect\label{FDT}
\end{equation}
where $\beta=1/T$ is the inverse temperature.

The fluctuation-dissipation theorem holds in the equilibrium regime,
but in the early  times of the dynamics one expects a breakdown of its
validity.  Mean Field studies \cite{CUKU,FM,BCKP} suggest the following
modification of the FDT (OFDR hereafter):
\begin{equation}
R(t_1,t_2)=\beta X(C(t_1,t_2)) \theta(t_1-t_2) \frac{\partial
C(t_1,t_2)}{\partial t_2} \; .
\protect\label{OFDR}
\end{equation}
where $X$  defines the violation of fluctuation-dissipation.
We can use the previous formula, eq.~(\ref{OFDR}), to relate the
observable quantities defined in eq.~(\ref{eq:auto}) and eq.~(\ref{res}).
In the linear response regime, the magnetization can be written as (we
report for completeness the formulas obtained for a Ising spin glass
in a magnetic field, so $m[h](t)\neq0$)
\begin{equation}
\begin{array}{l}
m[h+\hp](t) = m[h](t) \\\\
+\displaystyle\int_{-\infty}^t d
t^\prime ~ \left.\frac{\delta m[h'](t)}{\delta h'(t^\prime)}
\right|_{h'(t)=h(t)} \hp(t^\prime) + {\cal O}(\hp^2) 
\end{array}
\end{equation}
and so,
\begin{equation}
\Delta m[h,\hp](t) = \int_{-\infty}^t d t^\prime ~ R(t,t^\prime)
\hp(t^\prime) + {\cal O}(\hp^2) \; ,
\protect\label{LR}
\end{equation}
where we have defined $\Delta m[h,\hp](t) \equiv m[h+\hp](t)-m[h](t)$.
Eq.~(\ref{LR}) is just the linear-response theorem neglecting
higher orders in $\hp$.
By applying the OFDR we obtain the dependence of the
magnetization with time in a generic time-dependent magnetic field
(with a small strength), $\hp(t)$,
\begin{equation}
\Delta m[h,\hp](t) \simeq \beta \int_{-\infty}^t d t^\prime ~
X[C(t,t^\prime)] \frac{\partial C(t,t^\prime)}{\partial t^\prime}
\hp(t^\prime) \; .
\end{equation}

Next we let the system evolve with the unperturbed Hamiltonian 
until  $t=\tw$ and then we turn on the
perturbing magnetic field $\hp$ (hence, the system feels a magnetic
field $h+\hp$)\footnote{In the first numerical application of this
  method, Franz and Rieger~\cite{FRARIE} 
 chose another dependence of the magnetic
  field with time: $h(t)=h_0 \theta(\tw-t)$.}
.  Finally, with this choice of the magnetic field, we
can write 
\begin{equation}
\Delta m[h,\hp](t) \simeq \hp \beta \int_{\tw}^t d t^\prime ~
X[C(t,t^\prime)] \frac{\partial C(t,t^\prime)}{\partial t^\prime} 
\protect\label{mag_1}
\end{equation}
and 
\begin{equation}
\Delta m[h,\hp](t) \simeq \hp \beta \int_{C(t,\tw)}^1 d u ~
X[u] \; ,
\protect\label{mag_2}
\end{equation}
where we have used the fact that we are working with 
Ising spins.  In the equilibrium regime ($X=1$, as the fluctuation-dissipation
theorem  holds) we must obtain
\begin{equation}
\Delta m[h,\hp](t) \simeq \hp \beta (1 - C(t,\tw)) \; ,
\protect\label{mag_fdt}
\end{equation}
\ie $\Delta m[h,\hp](t)\, T/\hp$ is a linear function of $C(t,\tw)$
with slope $-1$. 

In the limit $t, \tw \to \infty$ with $C(t,\tw) = q$, one has that
$X(C) \to x(q)$, where $x(q)$ is given by
\begin{equation}
x(q)=\int_{q_{\mathrm{min}}}^q ~d q^\prime ~P(q^\prime)\; , 
\protect\label{x_q}
\end{equation}
where $P(q)$ is the equilibrium probability distribution of the
overlap with support $[q_{\mathrm{min}},q_{\mathrm{max}}]$. 
Obviously $x(q)$ is equal to 1 for all $q >
q_{\mathrm{max}}$, and we recover FDT for $C(t,\tw) >
q_{\mathrm{max}}$.  This link between the dynamical function $X(C)$
and the static one $x(q)$ has been already verified for finite
dimensional spin glasses~\cite{FDT}. The link has been analytically
proved for systems with the property of stochastic
stability~\cite{FRANZ}. 

We remark that we can use this formula to obtain
$q_{\mathrm{max}}$ as the point where the curve $\Delta m[h,\hp](t)$
against $C(t,\tw)$ leaves the line with slope $-\beta \hp$.

 For further use, we define 
\begin{equation}
S(C)\equiv \int_C^1 d q ~x(q)\; ,
\protect\label{s_c}
\end{equation}
or equivalently
\begin{equation}
P(q) =-\left.\frac{d^2 S(C)}{d^2 C}\right|_{C=q} \; .
\label{pq}
\end{equation}  
In the limit where $X \to x$ we can write eq.~(\ref{mag_2}) as
\begin{equation}
\frac{\Delta m[\hp](t)\;T}{\hp} \simeq S(C(t,\tw)) \ .
\protect\label{final}
\end{equation}

Looking at the relation between the correlation function and the
integrated response function for large $\tw$ we can thus obtain
$q_{\mathrm{max}}$, the maximum overlap with non-zero probability, as the point
where the function $S(C)$ becomes different from the function $1-C$.

From the function $S(C)$ we can get information on the overlap
distribution function $P(q)$, through eq.~(\ref{pq}).  Let us recall
which is the prediction for the $S(C)$ assuming the validity of each one of
the competing theories described in the introduction.  The droplet
model predicts $P(q)=\delta(q-{\hat q})$ and consequently\footnote{
In models with only one state, as the droplet model predicts for the
Ising spin glass in a magnetic field, 
the equilibrium time is finite irrespective of the value of the
volume of the system, hence, we can always thermalize any volume, and
so the asymptotic behavior, for waiting times larger than the
equilibration time, consists only of the straight line $1-C$. There
is no horizontal part.}

\begin{equation}
S(C) = \left\{
\begin{array}{cl}
1 - {\hat q} & {\mathrm{for}} \;\; C \le {\hat q} \ ,\\
1 - C & {\mathrm{for}} \;\; C > {\hat q} \; .
\end{array}
\right.
\protect\label{droplet}
\end{equation}

On the other hand the RSB prediction for the overlap
distribution\cite{MEPAVI}, 
$P(q) = (1-x_\mathrm{M}) \delta(q-q_{\mathrm{max}}) + x_\mathrm{M}
\delta(q-q_{\mathrm{min}}) + \tilde{p}(q)$ (where the support of
$\tilde{p}(q)$ belongs to the interval
$[q_{\mathrm{min}},q_{\mathrm{max}}]$, $q_{\mathrm{min}} \propto
h^{4/3}$ and $q_{\mathrm{max}}$ mainly depends on the temperature),
implies that
\begin{equation}
S(C) = \left\{
\begin{array}{cl}
S(0) & {\mathrm{for}} \;\; C \le q_{\mathrm{min}} \ ,\\
\tilde{s}(C) & {\mathrm{for}} \;\; q_{\mathrm{min}} < C \le q_{\mathrm{max}} \ ,\\
1 - C & {\mathrm{for}} \;\; C > q_{\mathrm{max}} \; ,
\end{array}
\right.
\protect\label{rsb}
\end{equation}
where $\tilde{s}(C)$ is a quite smooth and monotonically decreasing
function such that
\begin{equation}
\tilde{p}(q) = -\left.\frac{d^2\tilde{s}(C)}{d C^2}\right|_{C=q}\; .
\end{equation}

\begin{figure} 
\begin{center} 
\includegraphics[width=0.5\textwidth,height=0.3\textheight]{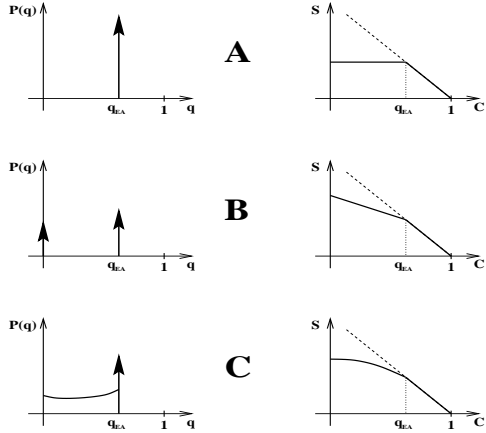} 
\caption{A possible model classification based on the function $S(C)$.
The big arrows represent delta functions. (A) corresponds to
droplet model, (B) to one step of replica symmetry breaking and (C)
to continuously broken replica symmetry (\eg Parisi solution of an
infinite dimensional Ising spin glass in absence of magnetic
field). Taken from reference~\cite{PRTRL_1999}.}
\label{fig:fdt_class}
\end{center} 
\end{figure}

In Figure \ref{fig:fdt_class} we show three possible behaviors of
the function $S(C)$ (and for the closely related function $P(q)$).

To finish this section we will recall an approximate scaling property
of the probability distribution of the overlap that was introduced by
Parisi and Toulouse (hereafter PaT)~\cite{ParisiToulouse}.  In
particular in Mean field the PaT hypothesis implies\footnote{The
goodness of this approximate Ansatz has been studied in 
reference~\cite{CRT_2003} in the Mean Field approximation. 
They find that none of the Parisi-Toulouse scaling hypotheses
about the $q(x)$ behavior hold, but some of them are only violated at
higher orders (taking as the parameter of the expansion the reduced
temperature).}

\begin{equation}
S(C) = \left\{
\begin{array}{cl}
1-C & {\mathrm{for}} \;\; C \ge q_{\mathrm{max}} \ ,\\
T \sqrt{1-C} & {\mathrm{for}} \;\; q_\mathrm{min} \le C 
\le q_{\mathrm{max}} \; .\\
\end{array}
\right.
\end{equation}

The result for $C \ge q_{\mathrm{max}}$ is general (and true for
finite dimension) and for $q_\mathrm{min} \le C \le q_{\mathrm{max}}$
we make the following Ansatz: $S(C)= A T (1-C)^B$ (in RSB $A=1$
and $B=1/2$).  If we substitute this Ansatz in eq.~(\ref{final})
we obtain the following scaling equation
\begin{equation}
\frac{m T}{h} T^{-\phi} =f\left( (1-C) T^{-\phi} \right)\; ,
\end{equation}
where $f$ is a scaling function and $\phi=1/(1-B)$ (in Mean
Field $\phi=2$). In order to be consistent, 
the scaling function should be  composed by a linear part
($x$) and by a power law part ($A x^B$).

In the rest of this section, we will discuss numerical simulations and
experiments. 

\subsection{Numerical Results}

\begin{figure} 
\begin{center} 
\includegraphics[width=0.5\textwidth,height=0.3\textheight]{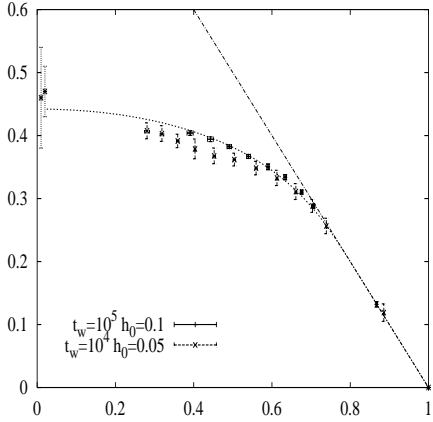} 
\caption{Off equilibrium fluctuation-dissipation relation. We plot $m(t,\tw)
  T/h$ versus the spin-spin correlation function $C(t,\tw)$ at
  $T\simeq 0.7 \Tc$. The lattice simulated was $L=64$ and we show
  two waiting times and two perturbing magnetic fields in order to
  control that linear response holds.
  The
  straight line with $-1$ slope is the equilibrium prediction. All the
  points on this line are pseudo-equilibrium points. We have marked
  the extrapolation to infinite time of the susceptibility with  the two
  leftmost points in the plot. Finally we have computed the $x(q)$
  function at equilibrium (from the numerical simulation, using
  parallel tempering~\cite{PARTEMP,ENZO}, of a $L=16$ lattice). 
  Taken from reference~\cite{FDT}.}
\label{fig:fdt}
\end{center} 
\end{figure}

 In Figure \ref{fig:fdt} we show the numerical points
obtained for two very large waiting times (in order to control that no
dependence on $\tw$ is found) and the prediction from the statics:
$x(q)$.  As a control we have computed the final point of the curve
(the $C=0$ point) extrapolating at infinite time the magnetization
using a power law fit.  The agreement is very good. Notice that the
asymptotic curve, which we can identify with the largest waiting time
in the figure, is not compatible with the prediction for the droplet model
(a horizontal part followed by the pseudo-equilibrium one)~\cite{FDT}.

From this figure, we can compute the Edwards-Anderson order parameter
($\qEA=q_\mathrm{max}$) 
as the point at which the numerical points depart from the
pseudo-equilibrium region (the straight line $1-C$). We can estimate
$\qEA\simeq 0.7$. If the droplet model holds the order parameter
should be $\qEA\simeq 0.55$ (in the DM, the asymptotic curve should be
a horizontal straight line in the region $[0,\qEA]$; the final
point $C=0$ is provided by the infinite time extrapolation of the
susceptibility, we can compute $\qEA=1-m T /h|_{\mathrm{asyn}})$.

\begin{figure} 
\begin{center} 
\includegraphics[width=0.5\textwidth,
height=0.3\textheight,angle=270]{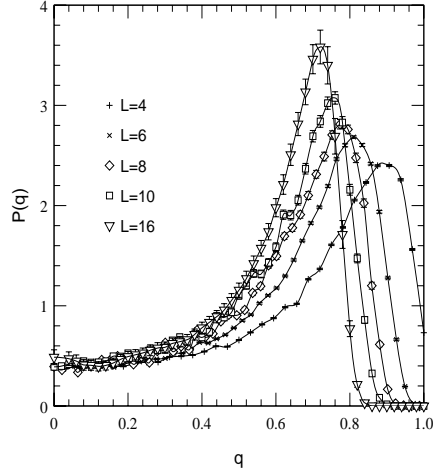}
\caption{Overlap probability distribution for $L=4,6,8,10$ and $16$ at
  $T\simeq 0.7 \Tc$. Taken from~\cite{COR1,INMAPARU}.}
\label{fig:pq}
\end{center} 
\end{figure}

\begin{figure} 
\begin{center} 
\includegraphics[width=0.5\textwidth,height=0.3
\textheight,angle=270]{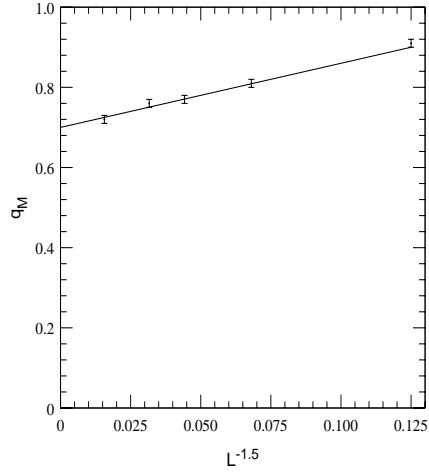}
\caption{Value of the overlap in which the probability distribution
  shows a maximum as a function of the lattice size at $T\simeq 0.7
  \Tc$. From \cite{INMAPARU}.}
\label{fig:qm}
\end{center} 
\end{figure}

We can test these possible values for $\qEA$. To do this we recall
equilibrium numerical simulation performed using parallel
tempering~\cite{PARTEMP,ENZO} in
a wide range of lattice sizes: $L=4,6,8,10$ and $16$. We plot in
Figure \ref{fig:pq} the equilibrium probability distribution of the
overlap $P(q)$. We can define $\qEA(L)$ as the value of $q$ in which
$P(q,L)$ shows a maximum. Furthermore we can analyze the dependence of
$\qEA(L)$ with $L$. The simplest dependence is a  power law:
\be
\qEA(L)=\qEA^\infty+\frac{a}{L^b} \; , 
\ee
where $a$ and $b$ are constants.
In Figure \ref{fig:qm}, we show $\qEA(L)$ versus $L^{-1.5}$ together with a
linear fit~\cite{INMAPARU}. 
Therefore, the data can be described with great accuracy
assuming a power law with a non zero value of $\qEA^\infty \simeq
0.7$. Finally, the data does not
support a power law fit with final value $\simeq 0.55$.\footnote{Incidentally, 
in the droplet model the probability of having an overlap
  different from the maximum one ($\qEA$) goes to zero as
  $L^{-\theta}$.  
  It is clear that data in Figure \ref{fig:pq} rule
  out this possibility. In particular, $P(0) \to \mathrm{const} \neq
  0$. The same conclusion is reached if one works with window overlaps
instead of the total ones~\cite{MAPARTRL}.} 
Therefore we have obtained two compatible estimates of $\qEA$
at $T=0.7$ using an off-equilibrium technique and an equilibrium one
and both results agree in the statistical error.

\begin{figure} 
\begin{center} 
\includegraphics[width=0.5\textwidth,height=0.3\textheight]{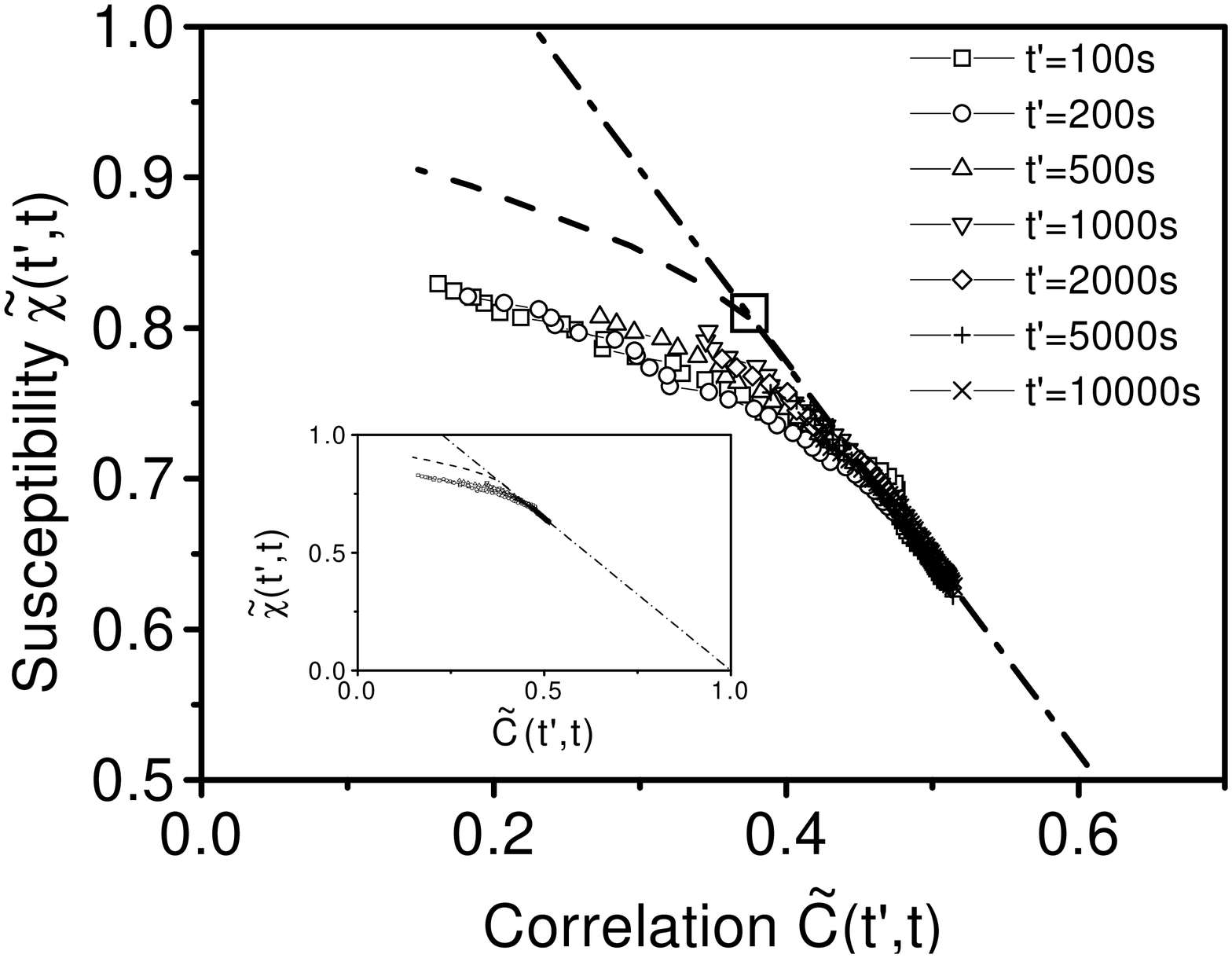}
\caption{Experimental determination of the function $X$ which induces
  the violation of fluctuation-dissipation. See the text for more
  details. Taken from reference~\cite{HO_2002}.}
\label{fig:fdt_exp}
\end{center} 
\end{figure}

This technique can be implemented in experiments. This has be done in
reference~\cite{HO_2002} by studying the
$\mathrm{Cd}\mathrm{Cr}_{1.7}\mathrm{In}_{0.3} \mathrm{S}_4$
insulating spin glass with $\Tg=16.2 \mathrm{K}$.  One measures the
response and the autocorrelation between the spins. The first part of
the work is not difficult, but the latter one has posed a challenge to
the experimentalists. We report in Figure \ref{fig:fdt_exp} the plot
of the violation of FDT. In contrast with what happens in numerical
simulations (where there is not a measurable dependence of the curves
with the waiting time for the larger times simulated. See Figures 3
and 4 of reference~\cite{SUE3} for a detailed study of the $L$ and
$\tw$ dependences), in the experiment a strong dependence  has been found 
for the reported curves with the waiting time, thus, an
extrapolation to large (infinite) waiting time is mandatory. This
extrapolation is the dashed line shown in the figure. Notice also the
dot-dashed line in Figure \ref{fig:fdt_exp} which corresponds to the
quasi-equilibrium regime. If one believes the extrapolation, the
figure supports heavily the RSB scenario and discards that of the 
droplet model.

\begin{figure} 
\begin{center} 
\includegraphics[width=0.5\textwidth,height=0.3\textheight,
angle=90]{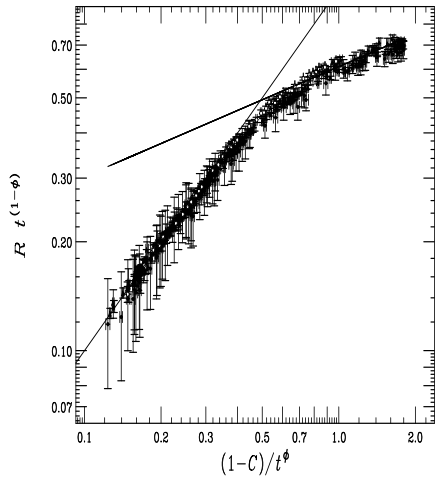}
\caption{PaT scaling for the three dimensional Ising spin glass
  ($h=0$). The plot has been built with $L$ and $\tw$ independent
  curves in order to check that we are in the asymptotic
  regime. Taken from reference~\cite{MPRTRL_1999}.}
\label{fig:pat}
\end{center} 
\end{figure}

Finally, we will end this section by showing a scaling analysis of
the off-equilibrium fluctuation-dissipation relations. This has been
done by using the PaT scaling which applies with great precision to
the equilibrium probability distribution, although it is not exact. In
Figure \ref{fig:pat} we report the scaling plot and it can be seen
that is a really good scaling (different magnetic fields, in order to
control linear response, waiting times, to check asymptoticity,
and temperatures)~\cite{MPRTRL_1999}.

Notice that the PaT scaling works for $L$ and $\tw$ independent
curves (see Figure \ref{fig:fdt} and reference~\cite{SUE3}). 
Two clear and distinctive regimes
can be seen in that figure. The first one correspond to the
quasi-equilibrium regime: in that part of the figure the behavior is
linear and thus it matches with the quasi-equilibrium regime $\Delta m T
/h = 1-C$. The second one corresponds to the aging regime: that part
of the plot can be parametrized with a power law with the $B$ exponent
introduced above.\footnote{
Following reference \cite{BarratBerthier} this kind of scaling  is
not enough to detect a RSB phase (they found in the two dimensional
Ising model ---with no phase transition at finite temperature--- a PaT
scaling for their OFDR). Nevertheless, in \cite{BarratBerthier} the
PaT scaling only works for points with the same waiting time, instead,
in the plot we have points computed with different waiting times. In
effect, we remark again, the scaling reported in Figure \ref{fig:pat} 
is $\tw$-independent (at least in
the numerical precision) which is a behavior completely different
from the two dimensional spin glass (paramagnetic phase).  For a
paramagnetic phase and very long waiting time (i.e. all the points lie
in the $1-C$ straight line) the PaT scaling plot should consist
in points over the linear part (quasi-equilibrium regime), and none
in the power law part (aging regime).}

\section{Memory and rejuvenation}

Maybe the  most striking features of spin glasses were found in
experiments where cycles in temperature were done. We are referring to
the so-called memory and rejuvenation  effects.

\begin{figure} 
\begin{center} 
\includegraphics[width=0.5\textwidth,height=0.3\textheight,
angle=0]{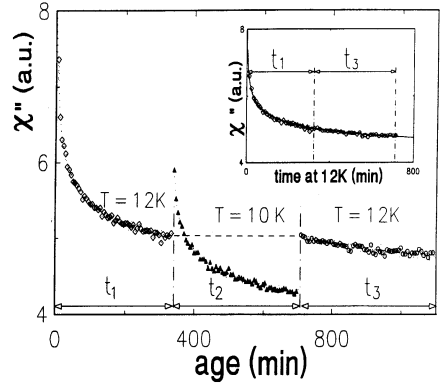}
\caption{Rejuvenation and memory in a real Ising spin glass. Out of
  phase susceptibility, $\chi^{\prime \prime}(\omega,t_a)$, of 
  $\mathrm{Cd}\mathrm{Cr}_{1.7}\mathrm{In}_{0.3} \mathrm{S}_4$, with
  critical temperature $\Tg=16.2 \mathrm{K}$, during a cycle in
  temperature. The frequency, $\omega$, is 0.01 Hz and $t_a$ is the
  time elapsed since the quench. The figure inset shows that, despite
  the strong relaxation at 10 K, both part at 12 K are in perfect
  continuation of each other. From~\cite{LHOV_1992}.}
\label{fig:memory_exp}
\end{center} 
\end{figure}

In Figure \ref{fig:memory_exp} we show an experimental plot reported
in Ref.~\cite{LHOV_1992}. In this experiment, a sudden quench from
high temperature is done to a temperature ($T=12$K) below the
critical one ($\Tg=16.2$ K). At this point the out of phase
susceptibility is recorded.  At a certain point of the experiment the
temperature is lowered again (in this case to $T=10$ K) and the out of
phase susceptibility is recorded again. As can be seen in Figure
\ref{fig:memory_exp} the out of phase susceptibility, measured at the
new temperature (10 K), starts from a higher value than the
susceptibility that the system had just before the quench from 12K to
10K. This is known as the rejuvenation of the system. When we cold a
system it behaves as if it was younger than before, \ie its out-of
phase susceptibility is higher than the one the system had at the
higher temperature just before the quench. In plain words, the system
at the new, lower temperature is farther from equilibrium than in the
last moments at the higher temperature. One can stay at the lower
temperature for a while and then restore the temperature of the system
to the original one (\ie we heat the system from 10 K to 12 K). In
Figure \ref{fig:memory_exp} we see that the system recovers the value
of the out of phase susceptibility that it had just before it was
cooled to 10 K. This phenomenon is known as memory effect. Notice in
the inset of Figure \ref{fig:memory_exp} how, despite the strong
relaxation produced at 10 K, the curves obtained in the higher
temperature $T=12$ K in two separated time intervals are in smooth
continuation. In reference~\cite{Jonason} the reader can see  good,
recent and detailed experimental studies of rejuvenation and memory
effects.

\begin{figure} 
\begin{center} 
\includegraphics[width=0.5\textwidth,height=0.3\textheight,
angle=0]{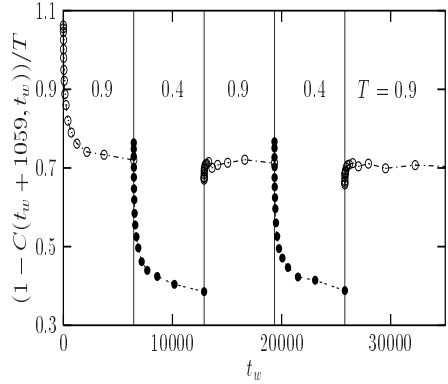}
\caption{Rejuvenation and memory in the numerical simulation of the
four dimensional Ising spin glass. Evolution of the ac correlation
function following the schedule: $T=\infty \to T_1=0.9 \to T_2=0.4
\to T_1 \to T_2 \to T_1$. Notice that the critical temperature for this model
is $\Tc=1.8$. Taken from \cite{BB_2002}.}
\label{fig:rejuvenation}
\end{center} 
\end{figure}

Berthier and Bouchaud~\cite{BB_2002} have recently obtained
rejuvenation and memory in the four dimensional Ising spin glass. We
reproduce in Figure \ref{fig:rejuvenation} their results. In three
dimensions  they have not seen these effects~\cite{BB_2002}.

As a numerical approximation to the  ac out of phase
susceptibility Berthier and Bouchaud~\cite{BB_2002} proposed to use:
\be
\chi(\omega,\tw)=\frac{1}{T} \left(1-C(\tw+\frac{1}{\omega},\tw)
\right) \;,
\ee
where $C(t,t^\prime)$ is the spin-spin dynamical correlation  defined
in eq. (\ref{eq:auto}). 

Following Berthier and Bouchaud we can try to explain memory and
rejuvenation effects in terms of the dynamical correlation
length~\cite{BB_2002}.

Let us first consider rejuvenation.  The system at the higher
temperature ($T_1$) thermalizes its so-called fast modes (\ie
$\xi(t,T_1) <<L$). When the system is frozen to a low temperature,
these fast modes that are just equilibrated at the higher temperature
are out of equilibrium in the new one, and so the system at the new
lower temperature, is younger than before ($\xi(t_1,T_2) <<
\xi(t_1,T_1)$, where $t_1$ is the time the system elapses in
$T=T_1$). This mechanism does not rely on the concept of chaos in spin
glasses~\cite{Jonason}\footnote{ Chaos, in this context, refers to
the sensitivity of equilibrium states in the ordered phase to
 small changes in the couplings or in temperature. Temperature
chaos postulates that typical equilibrium configuration at two
different temperatures $T_1$ and $T_2$ respectively, are strongly
correlated in a distance $l_0$ which depends on $\Delta
T=T_1-T_2$. For distances larger than $l_0$ the correlations between
these two typical configuration go to zero. Scaling arguments provides
$l_0 \sim |\Delta T|^{1/a}$, where $a=d_s/2-\theta$, $d_s$ is the
fractal dimension of the interface (see next section) and $\theta$ is
the usual droplet exponent. The change of the equilibrium states as we
change temperature explains rejuvenation.  In Mean Field the effect of
chaos in temperature is minimal~\cite{RIZZO}.  In numerical
simulations no clear chaos effects have been detected~\cite{Billoire},
but see~\cite{chaos_recent}; for a droplet interpretation,
see~\cite{chaos_droplet}.}.

The memory effect can be understood as follows. We have said that
rejuvenation involves the reorganization of small scales as compared
to the lengths involved in the aging at $T_1$, the higher
temperature. When we heat the system from $T_2$ to $T_1$, these small
scales ``almost instantaneously'' equilibrate at $T_1$, and the aging
restarts at $T_1$ at the same point. More quantitatively: the time
needed for the system  to recover its age at $T_1$ is given by
$\xi(t_\mathrm{memory},T_1)\simeq \xi(t_2,T_2)$, where $t_2$ is the
time elapsed in $T_2$. If $T_1-T_2$ is ``large'' then
$t_\mathrm{memory}<< t_2$.  So, in this interpretation, memory is
based in the existence of two, well separated, scales, while
rejuvenation is based in the reorganization of small scales.

\section{Spin glass at zero temperature}

In the last years a large amount of numerical work has been devoted
to numerical simulations at zero temperature. In particular  has
been studied the influence of perturbations in the ground state of the
system. We will study in the next two subsections, two way to perturb the
system and we will discuss the results in the light of the three scenarios
(RSB, DM and TNT).

\subsection{Changing the boundary conditions}

We will review in this section numerical simulations performed at zero
temperature in which ground states of the system are computed with
great accuracy (see reference~\cite{HART2}).

One interesting observable is the link overlap (we will study in the
next section its properties at finite temperature) defined by~\cite{MAPA}: 
\be
q_l(i,\bbox{\mu})=q(i) q(i+\bbox{\mu}) 
\ee 
where $q(i)=\sigma(i)
\tau(i)$ is the overlap, where $\sigma$/$\tau$ belong to a ground
state that has been computed using periodic/antiperiodic boundary
conditions respectively, and by $\bbox{\mu}$ we are denoting one of
the $d$ unitary vectors that can be defined in a $d$-dimensional
hypercubic lattice (and so $i+\bbox{\mu}$ is the point of the lattice
neighbor of $i$ in the direction provided by the vector
$\bbox{\mu}$).

Neglecting the points at the boundary, we can define the interface
between both ground state configurations as the region of  space in
which $q_l(i,\bbox{\mu})=-1$.
The probability to pick up such interface on a given random link, $l$,
is given by
\be
\rho=\frac{1}{2}(1-\overline{q_l}) \;,
\ee
where by $\overline{q_l}$ we denote the disorder expectation of
$q_l(i,\bbox{\mu})$ averaged over all sites, $i$, and directions,
$\bbox{\mu}$.

At this point one is  faced with three possibilities (there are an
additional fourth, but, we refer to the reader to
reference~\cite{MAPA} for a detailed explanation):

\begin{enumerate}

\item The interface is confined to a region of width $L ^z$, with
  $z<1$; inside of this region the interface could have overhangs. In
  this case $\rho$ goes to zero following a pure power law $L^{-\alpha}$,
  where $\alpha \ge 1-z$. \footnote{Roughly, the volume of the
  interface is $L^{d-1} \times L^z$. To obtain the probability we must
  divide the interface volume by the space volume, obtaining the relation
  between $\alpha$ and $z$.}

\item The wandering exponent $z$ is equal to 1 and the interface is a
  fractal object (not a multifractal) with fractal dimension,
  $d_s$. Then $\rho \sim L ^{-\alpha}$ with $\alpha=d-d_s$.

\item The exponent $\alpha$ is zero and the probability, $\rho$, goes
  to constant. Hence the interface is space filling. This last
  possibility is realized in the RSB scenario.

\end{enumerate}

Let us consider some examples. In a ferromagnet the ground state
obtained with antiperiodic boundary conditions (a.b.c.) should be
locally similar to that obtained with periodic boundary conditions
(p.b.c.), modulo an interface. As the interface, which is a flat
surface, has no measure, the link overlap between these two ground
states should be 1 in the large volume limit. In the droplet picture
of spin-glasses the discussion is similar. The only difference with
the previous example is that the interface is not necessarily flat rather  it
could be a corrugated surface (scenario 2). On the other hand, the RSB
of spin-glasses is different: The ground state obtained under
a.p.b. is expected to be locally similar to one of the low energy
states of the spectrum of the p.b.c. case. Thus, the link overlap
between these two ground states should tend in the large volume limit
to a constant value different from 1.

By computing ground states, Marinari and Parisi reported the following
values for the link overlap computed using four different
methods~\cite{MAPA}: $q_l=0.755(15)$, $0.80(6)$, $0.732(8)$ and
$0.722(5)$  (these figures have been extrapolated to $L\to\infty$). Putting
all four results together, we can finally quote \be q_l=0.79(7) .
\label{eq:T0}
\ee
Note that $q_l$ is  three standard deviations away from the droplet
prediction $q_l=1$.

\begin{figure} 
\begin{center} 
\includegraphics[width=0.5\textwidth,height=0.3\textheight,
angle=90]{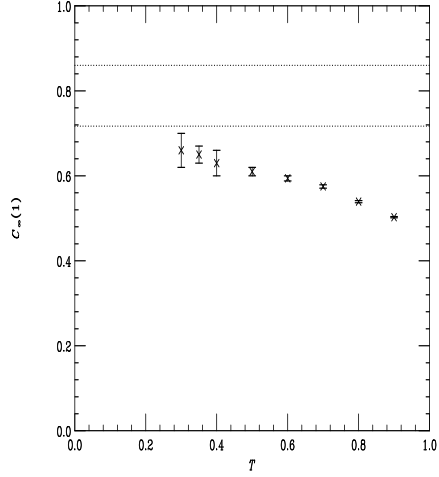}
\caption{Values extrapolated to infinite time of the correlation
function at distance $x=1$, $C_{\infty}(1,T)\equiv
C_{\mathrm{eq}}(x=1,T)$,  versus $T$. We have also marked by two horizontal
dotted lines the interval where the value computed in
reference~\cite{MAPA} lies, using $T=0$ ground state calculations.  The
consistency of the two results is clear. }
\label{fig:extra_T0}
\end{center} 
\end{figure}

We can try to recover this figure by performing numerical simulations 
at finite temperature and then try to extrapolate the data  to zero
temperature. This has been done in reference~\cite{MAPARL_2002}
 by noting that since
the overlap between the ground state computed with two different
boundary conditions (since the change of boundary conditions can be
regarded as a strong perturbation and the
corresponding ground states are far away) is expected to be very small
we can use the dynamical numerical simulations reported in the
previous section (in which the overlap remains almost zero all the run)
in order to obtain the value of the link overlap. It is easy to obtain
$q_l$ since it is nothing but $C_\mathrm{eq}(x=1)$ 
(see eq. (\ref{eq:pure_law})). In Figure
\ref{fig:extra_T0} we plot the values of $C_\mathrm{eq}(1)$ obtained
for different temperatures and we also mark the value obtained at zero
temperature, see eq. (\ref{eq:T0}). The consistency of  both sets
of data  is very good. We will come back to this issue at the end of
this section.

\subsection{Bulk perturbations}

Another way to perturb the system is to add a perturbation to the
couplings of the model~\cite{PAYO_2000,MP_PRL}. The new Hamiltonian reads:
\be
{\cal H}^\prime = {\cal H} + \frac{\epsilon}{N_B} \sum_{<i,j>} \sigma_i^0
\sigma_j^0 \sigma_i \sigma_j \; ,
\label{eq:bulk}
\ee
where $N_B=d L^d$ is the number of bonds, $\epsilon$ is the strength of the
perturbation and $\{\sigma^0\}$ is the ground state configuration computed
with no perturbation (\eg $\epsilon=0$). The sum $\sum_{<i,j>}$ is
extended over all pairs of nearest neighbors.

We can define the following link overlap
\be
q_l^{(\alpha, 0)}=\frac{1}{N_B} \sum_{<i,j>} \sigma_i^0
\sigma_j^0 \sigma_i^\alpha \sigma_j^\alpha \; .
\ee
One overlap can be defined in the usual way
\be
q=\frac{1}{L^d} \sum_{i} \sigma_i^0 \sigma_i^\alpha  \; .
\ee
Hence, we can write eq.(\ref{eq:bulk}) as
\be
{\cal H}^\prime = {\cal H} + \epsilon q_l^{(\alpha, 0)} \; ,
\ee
where we have labeled the configuration $\{\sigma\}$ by the index
$\alpha$.

One can show that the original ground state energy is shifted by an
amount  $\epsilon$ (since $q^{(0, 0)}=1$); 
the energy of any other state $\alpha$ (\eg a
low lying excitation of the spectrum) is shifted by an amount  $\epsilon
q_l^{(\alpha, 0)}$.

Let $\Delta E$ be the gap between the $\alpha$ state and the ground state
one in absence of perturbation. If $\epsilon> \epsilon
q_l + \Delta E$ the new ground state should be the $\alpha$ one. 

 One can compute the droplet prediction for the probability of
this event. In the DM  the probability to have an excitation of energy E
is given by\footnote{It has been assumed that the probability of an
excitation on the ground sate scales as $L^{\theta^\prime}$. In the
droplet picture the $\theta$ exponent is defined computing the scaling
of the difference of free energies from boundary condition changes. It
turns out that the droplet picture predicts $\theta=\theta^\prime$.}
\be 
P(E)=\frac{1}{L^{\theta^\prime}} f(\frac{E}{L^{\theta^\prime}}) \;, 
\label{eq:f}
\ee 
and so the probability to have an excitation with energy less
than $\epsilon(1-q_l)$ is given by \be P_<=\int_0^{\epsilon (1-q_l)}
dE~ P(E)=g(\frac{\epsilon}{L^\mu})\;, \ee 
where we have used that
$1-q_l\simeq L^{-(d-d_s)}$ and $g(u)\equiv\int_0^u ds~ f(s)$
\footnote{In the droplet picture there are only two 
thermodynamic states related by spin-flip reversal symmetry. The
overlap between a droplet and the ground state is one minus a term
which scales as the volume of the droplet ($L^d$) 
divided by the volume of the system ($L^d$). The link
overlap in this circumstance differs from 1 by a factor which is the
volume of the interface of the droplet ($L^{d_s}$) over the total
volume ($L^d$). We are assuming that the excitation is one droplet
with size proportional of that of the lattice
size~\cite{PAYO_2000}.}. Finally $\mu=\theta^\prime+d-d_s$.  
Now we can write the following scaling function for the
overlap and the link overlap (which are given by the product of the
probability of having a favorable droplet $P_<$, times the contribution
of this droplet to $1-\overline{q}$ and $1-\overline{q_l}$
respectively, see the previous footnote)
\begin{eqnarray}
1-\overline{q} &\propto&  g(\frac{\epsilon}{L^\mu})\;, \\
1-\overline{q_l} &\propto& \frac{1}{L^{(d-d_s)}} g(\frac{\epsilon}{L^\mu})  \;.
\end{eqnarray}   

We remark that by $\overline{q}$ and $\overline{q_l}$ we denote the
average of the overlap and the link overlap over the disorder,
respectively.  For small $\epsilon$, assuming $f(0) \neq 0$, we obtain
the following asymptotic formulas
\begin{eqnarray}
1-\overline{q} &\sim& \frac{1}{L^{d-d_s+\theta^\prime}}\;, \\
1-\overline{q_l} &\sim& \frac{1}{L^{2(d-d_s)+\theta^\prime}} \; .
\end{eqnarray}

Palassini and Young found for the three dimensional Gaussian spin
glass with periodic boundary conditions~\cite{PAYO_2000}
\be
\theta^\prime=0.02(3) \;,\;\; d-d_s=0.42(2) \;,\;\; \mu=0.42(3) \;, 
\ee
and in four dimensions~\cite{PAYO_2000}
\be
\theta^\prime=0.03(5) \;,\;\; d-d_s=0.23(2) \;.
\ee
This implies a non trivial probability distribution for the overlap but
a trivial one for the link overlap. 

However the RSB scenario cannot be ruled out by the data because it is
possible to fit $1-\overline{q}$ and $1-\overline{q_l}$ to a constant
plus scaling corrections, \ie
\be
1-\overline{q_l}=a+\frac{b}{L^c}
\ee
with $a=0.28(3)$ (\ie $q_l=0.72(3)$), $b$ and $c$  being positive constants. 
Indeed, these
are scaling-corrections that are compatible with the RSB prediction: 
$\theta^\prime=0$ and $d=d_s$.

On the other hand, the droplets prediction is
$\theta^\prime=\theta\simeq 0.2$ (in three dimensions) and
$d-d_s>0$. 

Krzalaka and Martin reached the same
conclusions~\cite{MARTIN}. Next, Houdayer, Krzakala and
Martin~\cite{HKM_2000} studied the topological properties of these
excitations finding sponge-like conformations (this provides a
geometrical picture~\cite{HOMA} for the RSB scenario) which costs
${\cal O}(1)$ in energy; finally they concluded that large finite size
effects should be presented in order to explain the data with the RSB
picture\footnote{ Houdayer, Krzakala and Martin provide three values
for the link overlap in their paper: $q_l=0.68$, 9.72 and 0.75
depending on the number of parameters used in their fits.}. Related
work by this group can be found in references~\cite{RELATED}.

Therefore they propose~\cite{MARTIN,PAYO_2000}, assuming the
absence of strong scaling corrections, an intermediate or mixed
scenario between droplet and RSB, the so called TNT picture (TNT for trivial
($q_l$), non trival ($q$)).

Yet, the controversy is not settled. In reference~\cite{MP_PRL}
Marinari and Parisi analyzing the data assuming RSB obtained
\be
1-q_l(q=0)=0.245(15)
\label{eq:T0bis}
\ee 
and by the study the correlation functions $1-q_l(q=0)=0.33(2)$. Here 
$q_l(q=0)$ denotes that the link overlap has been computed using only
configurations with mutual zero overlap. Moreover it has been found
that $q_l(q)$ depends quadratically on $q$, as found in infinite dimension.

In addition, reference~\cite{PLJY_2002} has obtained, by simulating the three
dimensional Gaussian Ising spin glass {\em but} with free boundary
conditions (in this reference the ground states were computed in an
exact way)
\be
1-\overline{q_l}^c=0.20(2)\;,
\ee
where the superscript $c$ denotes the average over those samples in
which the unperturbed and perturbed ground states are very different
(\ie the overlap is less than a given threshold value,
$q_\mathrm{max}=0.4$. Another cutoff $q_\mathrm{max}=0.2$ gives
essentially the same results). This value is in very good agreement
with that obtained by Marinari and Parisi and cited above in this
section (see eqs. (\ref{eq:T0}) and (\ref{eq:T0bis})). However, the authors
also found that if scaling corrections are allowed into the droplet
prediction, then an equally good fit is found, and one obtains
\be
\theta^\prime=0.19(6) \;,\;\; d-d_s=0.44(3) \;,\;\; \mu=0.63 \;.
\ee
Notice that $\theta^\prime=\theta\simeq 0.2$ as the droplet picture
predicts, but in contradiction with TNT.\footnote{Incidentally, the relation
  between the link overlap and the overlap has been study in 
this reference~\cite{PLJY_2002} obtaing $q_l=0.77(2)+0.27(3) q^2$,
according with RSB predictions. Moreover $q_l(q=0)=0.77(2)$.}

\section{(More on) The link overlap (at finite temperature)}

\begin{figure} 
\begin{center} 
\includegraphics[width=0.5\textwidth,height=0.3\textheight,
angle=0]{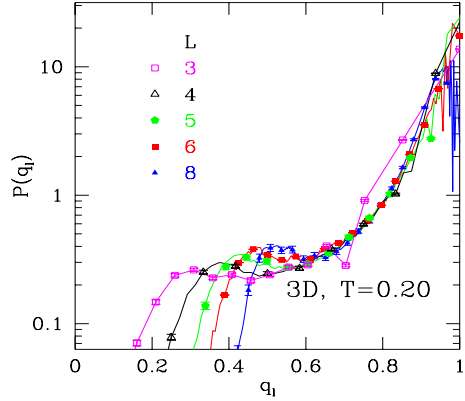}
\caption{$P(q_l)$ for the three dimensional Gaussian spin glass at
  $T=0.2$ for different lattice sizes with periodic boundary
  conditions. Taken from reference~\cite{KPY_2001}.}
\label{fig:pql}
\end{center} 
\end{figure}

The aim of this section is to study the properties of the link-overlap
$q_l$ {\em but} at finite temperature~\cite{KPY_2001,KY_2002}.  The
goal here is to characterize the probability distribution of the
overlap computing its variance.  It has been found that 
\be
\mathrm{var}(q_l) \sim L^{-\mu_l} \; .  
\ee 
It is possible to compute
the $\mu_l$ exponent as a function of $\theta^\prime$ and $d_s$, the
fractal dimension of an excitation.  If we assume that this variance
is dominated by the contribution of a single droplet of size $L$, then
this event occurs with probability $T/L^{\theta^\prime}$ (assuming a
constant density of states for these excitations, \ie $f(0) \neq 0$,
see eq. (\ref{eq:f})). We have seen that $1-q_l$ is proportional to $L
^{-(d-d_s)}$. The same holds true for $\delta q_l$. Hence, the variance
(which is the mean value of $\delta q_l^2$) is given by \be
\mathrm{var}(q_l) \sim \frac{T}{L^{\theta^\prime}} L^{-2( d-d_s)} \; ,
\ee and so $\mu_l=\theta^\prime+2 (d-d_s)$.  The link overlap
probability distribution obtained with numerical simulation of the
three dimensional Ising spin glass with periodic boundary conditions
is shown in Figure \ref{fig:pql}. In the droplet model this
probability distribution should shrink to a Dirac delta (zero variance); 
instead, in RSB $P(q_l)$ should have a  compact support (and
so non zero variance).

The extrapolation of the $\mu_l$ exponent to zero temperature gives a
value: $\mu_l=0.76(3)$. By assuming $\theta^\prime=0$ this implies
that $d-d_s=0.38(2)$, near  the value computed 
directly at $T=0$ using ground state computations: $d-d_s=0.42(2)$
(see preceding section). However, a
fit assuming RSB (\ie $\mu^\prime=0$) cannot be ruled out by the
numerical data~\cite{KPY_2001}.

\begin{figure} 
\begin{center} 
\includegraphics[width=0.5\textwidth,height=0.3\textheight,
angle=0]{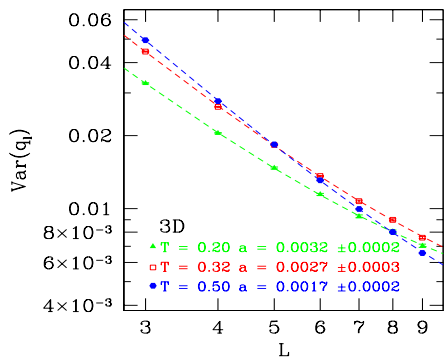}
\caption{Log-log plot of the variance of the link overlap as a function
  of the size (with free boundary conditions) for different
  temperatures. Taken from reference~\cite{KY_2002}.}
\label{fig:var}
\end{center} 
\end{figure}

In reference \cite{KY_2002} free boundary conditions (f.b.c) were
used.  With this f.b.c they try to discern
between a trivial behavior for the link overlap given by
$\mathrm{var}(q_l) \sim L^{-e}$ with a suitable $e$ exponent, and the  RSB
behavior given by $\mathrm{var}(q_l)= a + b L^{-c}$. They found a
finite value for $a$, for all the temperatures simulated, which
implies $d=d_s$ and that a pure law behavior ($L^{-e}$) is excluded by
the data. In Figure \ref{fig:var}  the variance of the link
overlap is plotted against the lattice sizes, in addition to the different fits
used (to $a+b/L^c$). The same analysis on the four dimensional Ising
spin glasses provides the same picture ($d=d_s$ and
$\theta^\prime=0$). However, let us end this section recalling
that it has been argued~\cite{PLJY_2002} that, in
principle, results obtained with f.b.c show larger finite-size  corrections
than results obtained with p.b.c. On the other hand,
f.b.c do not pose any restriction on the position of the domain
wall. Hence, it is not clear if the results reported for f.b.c
represent the asymptotic behavior and what are the optimal boundary
conditions for this kind of studies.

\section{Conclusions}

We have reviewed in detail (some) recent works on finite dimensional spin
glasses obtained with numerical simulations and in some cases, we have
done a direct comparison with experimental results. Moreover we have
tried to describe the numerical data with the competing three
theoretical pictures which try to describe the low temperature phase
of finite dimensional spin glasses.

Of course, as said in the introduction, we have omitted important
issues (we apologize), but we hope that this review will
clearly show which are the difficulties faced and the open/closed
problems regarding this interesting and active field of Statistical
Mechanics and Condensed Matter.

\section{Acknowledgments}

First of all, I warmly thank Enzo Marinari, Giorgio Parisi and
Federico Ricci-Tersenghi for a long and fruitful collaboration.
Moreover, I thank the people of the RTN group, with them I started to
do research in Statistical Mechanics and I still maintain an
interesting collaboration: Jos\'e Luis Alonso, Andr\'es Cruz, Luis
Antonio Fern\'andez, Antonio Mu\~noz, Alfonso Taranc\'on, Sergio
Jim\'enez, Jarda Pech, and V\'{\i}ctor Mart\'{\i}n-Mayor.  I thank
(again) V. Mart\'{\i}n-Mayor, Rodolfo Cuerno and Elka Korutcheva for a
critical reading of the manuscript.  I cannot forget Barbara Coluzzi,
David I\~niguez, Paola Ranieri, Marco Ferrero, Francesco Zuliani,
Daniel Stariolo, Remi Monasson, David Lancaster, Felix Ritort, Leticia
Cugliandolo, Jorge Kurchan, Jos\'e Carlos Ciria, Carlos L. Ullod and
Hector G. Ballesteros.  Finally I acknolwledge financial support from
the grants PB98-842, BFM2001-0718 (CICyT) and DYGLAGEMEM (European
Union).

\end{document}